\documentclass[onecolumn,superscriptaddress,amssymb,amsmath,nobibnotes,aps,showpacs,
prd,nofootinbib]{revtex4}%

\pdfoutput=1

\usepackage{graphicx}
\usepackage{epsfig}
\usepackage{bm}
\usepackage{amsfonts}
\usepackage{color}
\usepackage{amssymb}
\usepackage{float}
\graphicspath{ {images/} }

\newcommand{\bea}{\begin{eqnarray}}

\newcommand{\eea}{\end{eqnarray}}
\newcommand{\bma}{\begin{pmatrix}}
\newcommand{\ema}{\end{pmatrix}}
\newcommand{\be}{\begin{equation}}
\newcommand{\ee}{\end{equation}}
\newcommand{\beno}{\begin{equation*}}
\newcommand{\eeno}{\end{equation*}}

\newcommand{\pa}{\partial}

\begin{document}

\title{Superradiance Effect of a  Black Hole Immersed  in an Expanding Universe}

\author{Stella Kiorpelidi}
\email{skiorpel@central.ntua.gr}
\affiliation{Department of
Physics, National Technical University of Athens, Zografou Campus
GR 157 73, Athens, Greece}

\author{Konstantinos Ntrekis}
\email{drekosk@central.ntua.gr}
\affiliation{Department of
Physics, National Technical University of Athens, Zografou Campus
GR 157 73, Athens, Greece}

\author{Eleftherios Papantonopoulos}
\email{lpapa@central.ntua.gr}
\affiliation{Department of Physics,
National Technical University of Athens, Zografou Campus GR 157
73, Athens, Greece}

\begin{abstract}
We studied  the superradiance effect of a charge black hole  immersed  in an expanding Universe. We considered a test massive charged scalar field scattered off the horizon of the charge McVittie black hole. We carried out a detailed analysis of the electric energy extracted from the horizon of McVittie black hole in two different epochs of the expansion of the Universe, the dust dominated and radiation dominated epochs.
We found that we have the  superradiance effect in both epochs of the expansion of the Universe.  Our  study also provides evidence that we have extraction of energy from the horizon of the neutral McVittie black hole.
\end{abstract}

 \pacs{98.80.-k, 04.60.Bc, 04.50.Kd}

\maketitle

\section{Introduction}

The question of how  the dynamics of compact objects are affected by the cosmological expansion is a basic issue in  General Relativity (GR)
and it has been studied for a long time. One of the first attempts was carried out by McVittie \cite{McVittie} who studied how a local mass distribution in the form of a perfect fluid is affected by the Universe expansion. A spacetime metric was introduced having the information of the cosmic expansion and
which at small distances reproduces the  Schwarzschild spacetime while its large distance limit is a FRW spacetime. Since then there are many attempts to understand and give an answer to this question \cite{Price:2005iv,BalagueraAntolinez:2007mx,Carrera:2008pi,Carrera:2009ve,Nandra:2011ui}. These investigations showed that if the spacetime is described by a FRW metric then,  gravitating systems which are weekly coupled and of small size  compared to the Hubble radius
participate in the expansion, without any observational effect however, while for large scale structures with the size of  a large fraction of the Hubble radius, their dynamics and their properties are affected by the cosmological expansion  \cite{Busha:2003sz}.

A simple model was discussed in \cite{Price:2005iv} of a classical atom in
a de Sitter background, with a coupling between an electron and the central charge. In this simple model it was found that  if the  coupling is weak the
atom is following the Universe expansion, while if the
coupling is very strong the atom is only slightly perturbed and  does not expand. This "all or nothing" behaviour was criticized   in \cite{Faraoni:2007es}
in two respects, firstly that the cosmological
background was restricted to be de Sitter space and secondly on the simplified model of the classical atom.
It was found that this  "all or nothing" behaviour discussed in
\cite{Price:2005iv} persists only in the  de Sitter background and
 more general FRW backgrounds have different behaviour, allowing for  a local objects to  participate to the
cosmological expansion. Bound particle geodesics  in a McVittie spacetime were studied in \cite{Antoniou:2016obw}.
It was found that  relativistic effects tend to
destabilize bound systems leading to an earlier dissociation
 compared to the predictions in the
context of the Newtonian approximation.

In this work we study the radiation of compact objects in an expanding Universe. In particular we investigate the  superradiance effect of a charge black hole  immersed  in an expanding Universe. We will consider a charged scalar field scattered off the horizon of the charge McVittie black hole \cite{Gao:2004cr,Faraoni:2014nba}.
 The superradiance effect is a result of extracting charge and electric energy from a  charged black hole \cite{Bekenstein}.  The superradiant scattering
of charged scalar waves in this  regime  may lead to an
instability of the black hole spacetime.

 We will show that as the Universe expands  we  have extraction of charge and electric energy from the horizon of the charge McVittie black hole. For two epochs of the expansion of Universe, the dust dominated and radiation dominated epochs we will perform a detailed analysis of the superradiance effect for a wide  range of values of the frequency of the scattered wave off the horizon of the charge McVittie black hole.

As it is well known we have the superradiance effect for a rotating  black hole and for a static charge black hole. We will provide evidence that
for a neutral McVittie black hole we also have the superradiance effect.  For a range of frequencies of a  scattered wave off the horizon of the neutral  McVittie black hole we  find that there is extraction of energy from its horizon.

\section{The Charged McVittie metric}

A charged  McVittie solution was constructed in \cite{Gao:2004cr} which can be considered as a metric of the
Reissner-Nordstr$\ddot{o}$m black hole immersed in a FRW expanding Universe. Assuming a slow evolution of the Universe and using the formalism for
computing the mass and charge of stationary
spacetime, it was  found that both the mass and charge of the black
hole decrease with the expansion of the Universe and increase with
the contraction of the Universe. It was also found that the two typical
scales, the time-like surface and the event horizon of the black
hole both shrink with the expansion of the Universe and expand
with the contraction of the universe.
 A detailed analysis
  of the location of the apparent horizons and a study of their dynamics is presented in \cite{Faraoni:2014nba}.
Considering various cases for the scale factor it was found that for the charged McVittie spacetime
in the explored  parameter range $|Q| \leq m$, a cosmological apparent
horizon and a black hole apparent horizon exist. We review in this Section the basic features of the charged McVitte metric.

Consider the action
\be
S= \int d^4 x \; \sqrt{-g}\left[\frac{R}{2} + \alpha F^{\mu\nu} F_{\mu\nu} + \mathcal{L}_{\text{fluid}} -  \frac{\eta}{2}(D_\mu \phi)^* (D^\mu \phi)  - \frac{\mu^2}{2}  \phi \phi^* \right]~, \label{action}
\ee
where $D_\mu = \nabla_\mu - i q A_\mu$ and $F_{\mu\nu} = \nabla_{\mu} A_\nu - \nabla_\nu A_\mu$. Varying the above action the Einstein equations become
\be
G_{\mu\nu} = T_{\mu\nu}^{(EM)} + T_{\mu\nu}^{\text{fluid}} + T_{\mu\nu}^{\phi}~,
\ee
where
\be
 T_{\mu\nu}^{(EM)} = -4 \alpha F^\kappa_\mu F_{\nu\kappa} + \alpha g_{\mu\nu} F_{\kappa\lambda}F^{\kappa\lambda}~,
\ee
and
\begin{align*}
T_{\mu\nu}^{\phi} = -\frac{\mu^2}{2} g_{\mu\nu}{|\phi|^2} + \frac{\eta}{2} \Big(-g_{\mu\nu} \nabla_\kappa \phi^* \nabla^\kappa \phi + \nabla_\mu\phi^* \nabla_\nu \phi + \nabla_\mu\phi \nabla^\nu \phi^* - 2 q^2 A_\mu A_\nu |\phi|^2 \nonumber\\  - q^2 g_{\mu\nu} A_\kappa A^\kappa  |\phi|^2 - i q A^\kappa g_{\mu\nu}\left(\phi^* \nabla_\kappa \phi + \phi \nabla_\kappa \phi^* \right)
\Big)~.
\end{align*}
Assuming that the scalar field $\phi$ does not backreact  to the metric $($that is, keeping linear terms of $\phi)$ we end up with the Einstein equations
\begin{align}
G_{\mu\nu} &= T_{\mu\nu}^{(EM)} + T_{\mu\nu}^{\text{fluid}}~.\\
\end{align}
A solution to the above equations with the electromagnetic potential as
\begin{equation}
A_\mu=\left\lbrace-\dfrac{Q}{R},0,0,0\right\rbrace~,
\end{equation}
is the charged McVittie metric
\begin{align} \label{metricR}
ds^2=-\left(1-\dfrac{2M}{R}+\dfrac{Q^2}{R^2}-R^2H^2\right) dt^2+\dfrac{R^2}{Q^2-2M R+R^2}dR^2
-\dfrac{2RH}{\sqrt{\dfrac{Q^2-2 M R+R^2}{R^2}}} dtdR+R^2d\Omega^2~,
\end{align}
where $R$ is the areal radius
\be \label{radius}
R(t,r)= r a(t) \left(\left(\dfrac{m}{2 r a(t)}+1\right)^2-\dfrac{Q^2}{4 r^2 a(t)^2}\right) \quad\Rightarrow\quad 2 a(t) r =  \sqrt{-2 m R+Q^2+R^2}-m+R~,
\ee
and the tensor of the fluid is ${T^\mu_\nu{}}^\text{fluid} =\left( \rho(t)+p(t,r)\right)U^\mu U_\nu + p(t,r)\delta^\mu_\nu$ with $U_\mu = \left(\dfrac{\sqrt{Q^2-2 M R+R^2}}{R},0,0,0\right)$.
The location of the apparent horizons is the solution of
\begin{equation} \label{apH}
\nabla_\mu R \nabla^\mu R = 0 \quad \Leftrightarrow \quad Q^2-2MR+R^2-R^4H^2(t)=0~.
\end{equation}
A detailed derivation of the charged McVittie solution is given in the Appendix A.

\section{The Dynamics of a Scalar Field in the Reissner-Nordstr$\ddot{O}$m - de Sitter - Kottler Spacetime}
\label{rnostsup}

The Reissner-Nordstr$\ddot{o}$m-de Sitter-Kottler (RNdSK) solution is representing a charged McVittie black hole of constant expansion rate $H$ which plays the role of a cosmological constant. The line element reads
\begin{equation}
ds^2=-\left(1- \dfrac{2 M}{R} - \dfrac{Q^2}{R^2} - H^2 R^2\right) dT^2 + \left(1- \dfrac{2 M}{R} - \dfrac{Q^2}{R^2} - H^2 R^2\right)^{-1} dR^2 +
 R^2 d\Omega^2~,  \label{rndeSitter}
\end{equation}
and it comes from the charged McVittie  metric \eqref{metricR} with the additional time coordinate transformation defined by
\begin{equation}
dT=dt+H\left(1-\dfrac{2M}{R}+\dfrac{Q^2}{R^2}\right)^{-1/2} \left(1-\dfrac{2M}{R}+\dfrac{Q^2}{R^2}-R^2H^2\right)^{-1} dR~.
\end{equation}
For an expanding universe with $\rho=-p = \Lambda$, where $\Lambda$ is the cosmological constant, and a scale factor $a(t)\sim e^{\sqrt{\frac{\Lambda}{3}}t}$, the Hubble parameter is constant $H=\sqrt{\Lambda /3}$.

\begin{figure}[h!]
\centering
\includegraphics[width=0.5\textwidth]{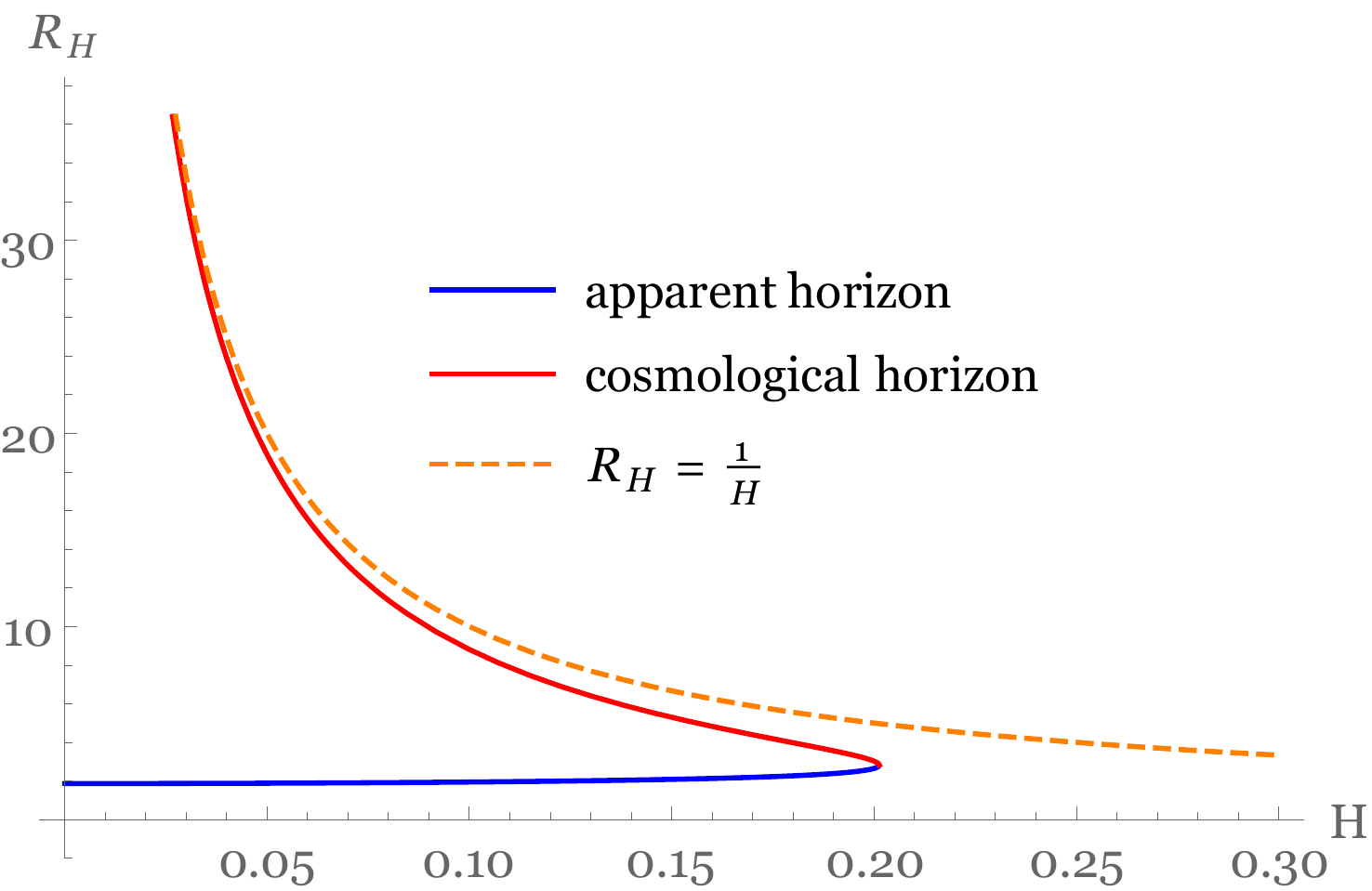}
\caption{The areal radius $R$ of the apparent horizons of the charged Reissner-Nordstr$\ddot{o}$m-de Sitter-Kottler spacetime versus cosmological constant $H$ for $M=1$ and $Q=1/2$. The blue and red  lines are the apparent and cosmological horizons respectively. The orange dashed line is the location of the corrected Hubble radius $R=1/H$. }\label{Qmi}
\end{figure}

The locations of the apparent horizons of the metric (\ref{rndeSitter}) are defined by the positive roots of the equation
\begin{equation}
1- \dfrac{2 M}{R} - \dfrac{Q^2}{R^2} - H^2
R^2=0~.
\end{equation}
A specific example of the solutions of the above equation is shown in Fig.~\ref{Qmi}.\\
\\
Since $M$ and $H$ are both necessarily positive (if we consider expanding universe) one of the horizons is negative and therefore unphysical. Concerning the two horizons, we refer to the blue line as the black hole apparent horizon, since it reduces simply to the Reissner-Nordstr$\ddot{o}$m horizon at $M+\sqrt{M^2-Q^2}$ if there is no background expansion $H \to 0$, while we refer to the red line as the cosmological apparent horizon, since it reduces to the static de-Sitter horizon at $1/H$ if there is no mass and charge present.

We now examine if the massive charged scalar field $\phi$ of our theory is scattered from the RNdSK black hole. The dynamics of $\phi$ are described by the Klein Gordon equation
\begin{equation}\label{KGR}
\left[\left(\nabla^\nu - i q A^\nu \right)\left(\nabla_\nu - i q A_\nu \right)-\mu^2\right] \phi=0~,
\end{equation}
where $A_\nu=- \delta^0_\nu Q/R$ is the electromagnetic potential of the black hole and $q$ and $\mu$ are the charge and mass of the scalar field, respectively. Following the procedure presented in \cite{Hod:2013eea,Hod:2013nn},  we decompose the scalar field as
\begin{equation}
\phi(T,R,\theta,\varphi) = e^{i m \varphi} S_{lm} (\theta) R_{lm}(R) e^{-i \omega T}~, \label{wavefunction}
\end{equation}
where $\omega$ is the conserved frequency of the mode, $l$ is the spherical harmonic index, and $m$ is the azimuthial harmonic index with $-l \leq m \leq l$ . The sign of $\omega_I$ determines whether the solution is decaying $\omega_I<0$ or growing $\omega_I>0$ in time.

Substituting (\ref{wavefunction}) in the Klein-Gordon equation (\ref{KGR}) we obtain the radial Klein Gordon equation
\begin{equation}\label{kgD}
\Delta \dfrac{d}{dR} \left(\Delta \dfrac{d R_{lm}}{dR} \right) + U R=0~,
\end{equation}
where
\begin{equation}
\Delta \equiv R^2 - 2 M R + Q^2-H^2 R^4~,
\end{equation}
and
\begin{equation}
U\equiv \left(\omega R-qQ \right)^2- \Delta \left[\mu^2 r^2 +l (l+1)\right]~.
\end{equation}
The equation $\Delta=0$ gives the black hole apparent horizon and the black hole cosmological apparent horizon. We are only interested in solutions of the equation \eqref{kgD} with the boundary conditions of purely ingoing waves at the black hole apparent horizon and a decaying (bounded) solution at spatial infinity. Using the transformation $dR^* = \left(R^2 / \Delta \right) dr$ the radial equation can take the form of a Schrodinger-like equation
\begin{equation}
\dfrac{d^2  R_{lm}(R^*)}{dR^*}+ \dfrac{U}{R^4} R_{lm}(R^*)=0 \label{kgshcrs}~.
\end{equation}
At the apparent horizon we have $\Delta=0$ and equation (\ref{kgshcrs}) has a solution
\begin{equation}
R_{lm}(R^*) \sim \mathcal{T}\; e^{- i \left(\omega -q Q/R_H \right) R^*}  \quad  \text{as} \quad R\to R_H \left(R^* \to - \infty\right)~, \label{horsol}
\end{equation}
while for $R^* \to + \infty$ we consider the case which $\mu = \sqrt{2} H $ so Eq.~(\ref{kgshcrs}) can take the form
\begin{equation}
\dfrac{d^2 R_{lm}}{dR^{*2}} +\left(\omega^2+H^2 l(l+1) \right) R_{lm}=0~,
\end{equation}
which admits the solution
\begin{equation}
R_{lm}(R^*) \sim   \mathcal{R}\; e^{i\sqrt{{H}^2 l (l+1)+\omega ^2} R^*} +  \mathcal{I}\; e^{-i\sqrt{{H}^2 l (l+1)+\omega ^2} R^*}  \quad  \text{as} \quad R\to \infty\; (R^*\to \infty)\label{infsol}~,
\end{equation}
where these boundary conditions correspond to an incident wave of amplitude $\mathcal{I}$
from spatial infinity giving rise to a reflected wave of amplitude $\mathcal{R}$ and a transmitted wave of amplitude $\mathcal{T}$ at the horizon (for a review see  \cite{Brito:2015oca}). For the frequencies in the superradiant regime
 the boundary condition (\ref{horsol}) describes an
outgoing flux of energy and charge from the charged black hole
\cite{Bekenstein,PressTeu1}. As it was shown in \cite{Damour:1976kh} these boundary conditions
lead to  a
discrete set of resonances $\{\omega_n\}$ which correspond to the
bound states of the charged massive field.

There are various ways to find the exact superradiance condition. One way   is to compute the greybody factor and the reflection coefficients in a  static spacetime. Then  if the greybody factor is negative or the reflection coefficients is greater than 1 \cite{Benone:2015bst} then the scalar waves can be superradiantly amplified. The easiest way is by
 using the Wronskian demanding the reflected amplitude to be bigger than the amplitude of the incident wave
\begin{equation}
|\mathcal{R}|^2 = |\mathcal{I}|^2 - \frac{\omega-qQ/ R_H}{\sqrt{H^2 l (l+1)+\omega ^2}}  |\mathcal{T}|^2~.
\end{equation}
For frequencies in the superradiant regime
\begin{equation}
\omega < qQ/R_H~, \label{bekenstein}
\end{equation}
which is known as the Bekenstein condition \cite{Bekenstein}, the boundary condition (\ref{horsol}) describes an outgoing flux of energy and charge from the charged de-Sitter black hole.

In \cite{Zhu:2014sya} small charged scalar field perturbations in the vicinity of a (3+1)-dimensional Reissner-Nordstr$\ddot{o}$m  black hole in a de Sitter background were studied. An instability was found of the Reissner-Nordstr$\ddot{o}$m black hole and it was argued  that this instability was due to superradiance satisfying the Bekenstein condition \eqref{bekenstein}.

\section{The dynamics of a scalar field in an expanding McVittie spacetime}
\label{mcVitsup}

In this  Section we will study a massive charged scalar field scattered off the charged McVittie black hole in an expanding Universe. In particular we will study two epochs of the Universe expansion, the radiation and dust dominated epochs.  Because the background spacetime is time dependent the scalar field should also be time dependent. Therefore, we consider the following decomposition of the scalar field
 $\phi(t,R,\theta,\varphi)$ as
\be
\phi(t,R,\theta,\varphi) = R_{lm}(t,R) Y_{lm}(\theta,\phi)~. \label{wave1}
\ee
The Klein-Gordon equation \eqref{KGR} in the background of the charged McVittie metric \eqref{metricR} takes the form
\begin{align}
-&\frac{\pa^2 R_{lm} (t,R)}{\pa R^2}\frac{Q^2-2MR+R^2-R^4H^2}{R^2}  + \frac{\pa^2 R_{lm} (t,R)}{\pa t^2} \frac{R^2}{Q^2-2MR+R^2}  +  \frac{\pa^2  R_{lm} (t,R)}{\pa t \pa R} \frac{2 R H}{\sqrt{\frac{Q^2-2MR+R^2}{R^2}}}  \nonumber \\
+ &\frac{\pa  R_{lm} (t,R)}{\pa R} \left(\frac{2(M-R)}{R^2} +4RH^2+\frac{R\dot{H}}{\sqrt{\frac{Q^2-2MR+R^2}{R^2}}}+ \frac{2 i qQH}{\sqrt{\frac{Q^2-2MR+R^2}{R^2}}}\right)  \nonumber\\
+&\frac{\pa  R_{lm} (t,R)}{\pa t}  \left(\frac{2 i qQR}{Q^2-2MR+R^2}+ H \frac{4Q^2-7MR+3R^2}{\sqrt{\frac{Q^2-2MR+R^2}{R^2}} \left(Q^2-2MR+R^2\right)}\right)\nonumber\\
+ &R_{lm}(t,R)\left(\frac{i q Q H(t) \left(R (2 R-5 M)+3 Q^2\right)}{\left(Q^2-2 M R+R^2\right)^{3/2}}+\mu ^2+\frac{ l (l+1)}{R^2}-\frac{q^2 Q^2}{Q^2-2 M R+R^2}\right)= 0~. \label{fullKGR}
\end{align}
For large $R$ the above equation takes the asymptotic form
\begin{align}
 &R^2H^2 \frac{\pa^2 R_{lm} (t,R)}{\pa R^2} +\frac{\pa^2 R_{lm} (t,R)}{\pa t^2} + 2RH  \frac{\pa^2  R_{lm} (t,R)}{\pa t \pa R}    \nonumber\\ +
 &\left(4RH^2+R\dot{H} \right) \frac{\pa  R_{lm} (t,R)}{\pa R}
+ 3H \frac{\pa  R_{lm} (t,R)}{\pa t} +
\mu^2 R_{lm} (t,R) = 0~.\label{KGLar}
\end{align}
The  derivative of the Hubble parameter $\dot{H}$ can be found from the pressure relation given in the Appendix in  equation (\ref{pieshH}).

\subsection{Dust dominated Universe}

We will study first the Klein-Gordon equation \eqref{fullKGR} for a dust dominated Universe. In this case $p(t,R) = 0$ and from equation (\ref{pieshH}) we get
\begin{equation}\label{Hdot}
\dot{H}(t)=-\dfrac{3}{2} H^2(t) \dfrac{\sqrt{Q^2-2 M R+R^2}}{R}~.
\end{equation}
This expression is expected since for large $R$ $($large r$)$ the charged McVittie metric \eqref{mcmetric} reduces to the FRW metric. Since $A_\mu = \left(-\dfrac{Q}{R},0,0,0 \right)$,  for large $R$ we get $D_\mu \to \nabla_\mu$ so the term with charge couplings disappears. Equation \eqref{KGLar} admits a solution of separate variables of the form
\begin{equation}\label{solLargeR}
R_{lm}(t,R) = R^\alpha T(t)~,
\end{equation}
where $\alpha$ is a complex number. We can also write $R^\alpha = R^{\Re(a)}\; e^{i \Im(\alpha) \ln(R)}$ from which is easy to see that for $\Re(\alpha) < 0 $ the solution goes to zero for $R\to \infty$. Concerning the function of time we find
\begin{equation} \label{eqT}
T(t) \left(\alpha  \dot{H}(t)+\alpha  (\alpha +3) H(t)^2+\mu ^2\right)+(2 \alpha -3) H(t) \dot{T}(t)+\ddot{T}(t) = 0~.
\end{equation}
For a dust dominated Universe, for large $R$, equation \eqref{Hdot} becomes
$
\dot{H}(t) = - \frac{3}{2} H^2(t)$,
with the solution $H(t) = \frac{2}{3 t}$. Substituting this in $\eqref{eqT}$ we find
\begin{equation}
T(t)=e^{- i \mu t} t^{-1-\dfrac{2 \alpha}{3}} C_1 - i \dfrac{1}{2 \mu} e^{i \mu t} t^{-1-\dfrac{2 \alpha}{3}} C_2~.
\end{equation}
So the asymptotic solution of the scalar field is
\begin{equation}
\phi^{R \to \infty} (t,R,\theta,\phi) = R^\alpha \left(e^{- i \mu t} t^{-1-\dfrac{2 \alpha}{3}} C_1 - i \dfrac{1}{2 \mu} e^{i \mu t} t^{-1-\dfrac{2 \alpha}{3}} C_2 \right) Y_{lm}(\theta ,\phi)~.
\end{equation}

At the opposite limit ($R=R_H$) substituting \eqref{apH} and \eqref{Hdotradiation} in the Klein-Gordon equation \eqref{fullKGR}
we find
\begin{align}
&R_{lm}(t,R)\left(\frac{\left((l+1) l+\mu ^2 R^2\right) \left(R (R-2 M)+Q^2\right)}{R^4}+\frac{i q Q \left(3 Q^2-5 M R+R^2 (2+i q Q)\right)}{R^4}\right)\nonumber\\
+ &\frac{\partial R_{lm}(t,R)}{\partial R}\frac{\left(R (R-2 M)+Q^2\right) \left(R (R-6 M)+4 i q Q R^2+5 Q^2\right)}{2 R^5}  \nonumber \\
+& \frac{\partial R_{lm}(t,R)}{\partial t} \frac{2 i q Q R^2+4 Q^2-7 M R+3 R^2}{R^3} \nonumber\\
+& \frac{\partial^2 R_{lm}(t,R)}{\partial R\partial t} \frac{2 \left(R (R-2 M)+Q^2\right)}{R^2} +  \frac{\partial^2 R_{lm}(t,R)}{\partial t^2} = 0~. \label{fullKGRH}
\end{align}
This equation admits a solution of the form
\begin{equation} \label{radialHOR}
R_{lm}(t,R)=e^{-i \omega t} e^{ \left(A+ i B\right) R }~,
\end{equation}
where
\begin{align}
A=&\dfrac{1}{\left(Q^2-2MR_H+R_H^2\right) \left(\left(5Q^2-6MR_H+R_H^2\right)^2+2 R_H\left(\omega R_H-q Q\right)^2\right)}  \nonumber \\
\cdot & \left(- 2 R_H^3 \left(\omega R_H-q Q\right)^2 \left(11Q^2-22MR_H+13R_H^2\right) -8qQR_H^3  \left(\omega R_H-q Q\right)\left(Q^2-2MR_H+R_H^2\right)\right. \nonumber \\
& -2 R_H \left(Q^2-2MR_H+R_H^2\right)\left(l(l+1)+\mu^2 R_H^2\right)\left(5Q^2-6MR_H+R_H^2\right) \Big)~,\\
B=&\dfrac{1}{\left(Q^2-2MR_H+R_H^2\right) \left(\left(5Q^2-6MR_H+R_H^2\right)^2+2 R_H\left(\omega R_H-q Q\right)^2\right)}  \nonumber \\
\cdot & \Bigg( 4R_H^4 \left(\omega R_H-q Q\right)^3   +qQ\left(l(l+1)+\mu^2 R_H^2\right)\left(5Q^2-6MR_H+R_H^2\right)\nonumber \\
&+ \Big(\left(5Q^2-6MR_H+R_H^2\right)\left(4Q^2-7MR_H+3 R_H^2\right) -4 R_H^2 \left(l(l+1)+\mu^2 R_H^2\right)\left(Q^2-2MR_H+R_H^2\right)\Big) \left(\omega R_H-q Q\right) \Bigg) \label{B}~,
\end{align}
and $\omega$ is the frequency of the mode.

As we discussed in Section \ref{rnostsup}, the solution (\ref{radialHOR}) corresponds to the boundary condition (\ref{horsol}) of the static RNdSK black hole and if $B>0$ it describes
 an outgoing flux of energy and charge from the charged McVittie black hole.

\begin{figure}[h!]
\centering
\includegraphics[scale=0.5]{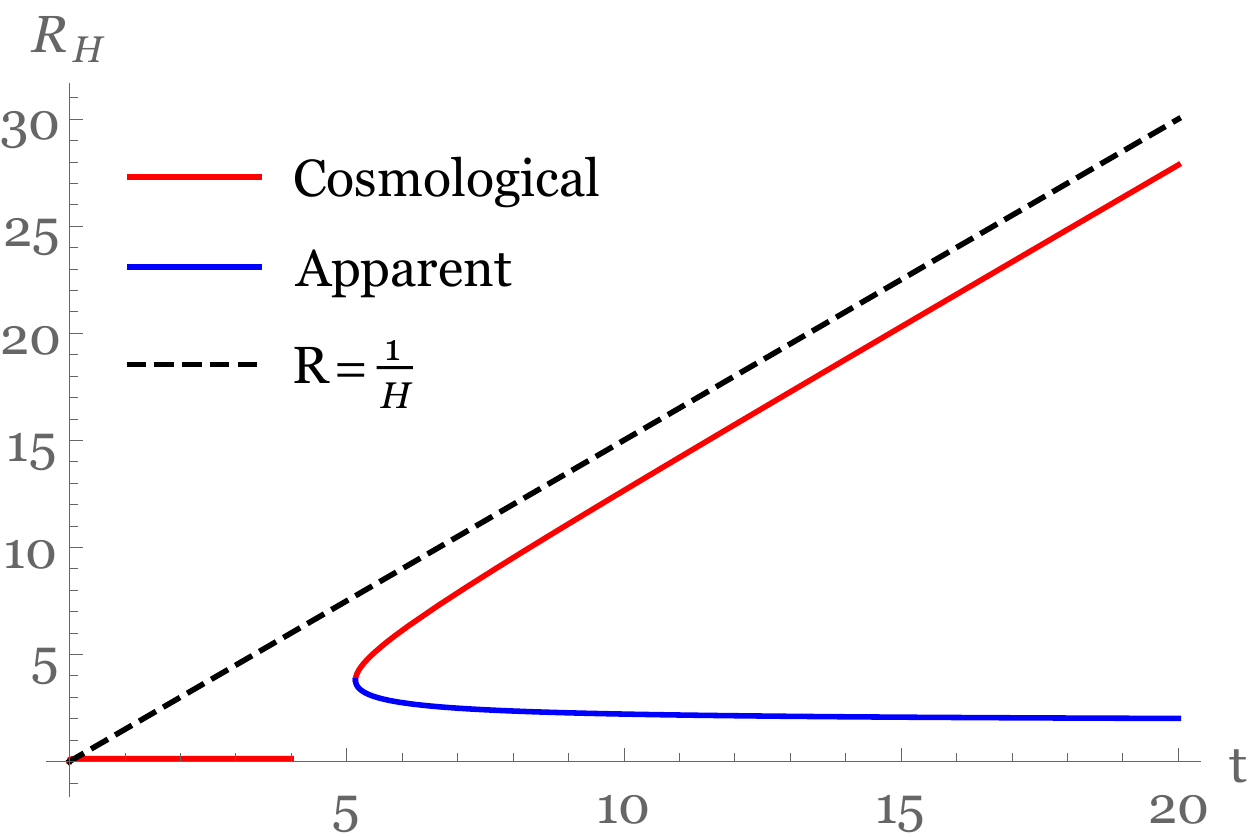}
\caption{Apparent and cosmological horizon for $Q=1/2$, $M=1$}\label{plotR_HQmisohorizons}
\end{figure}

After a critical time $t^*$ the charged McVittie has two horizons, an apparent horizon and a cosmological. The scattering of a scalar field takes place at the apparent horizon, therefore this critical time can give us an  upper limit to the size of black hole where the superradiance phenomenon occurs. As $t \to \infty$, we see that the apparent horizon shrinks at $M+\sqrt{M^2-Q^2}$ which gives us a lower limit to the $R_H$.
Comparing Fig.~\ref{plotR_HQmisohorizons} with the Fig.~\ref{Qmi} showing the horizons of the  RNdSK black hole  can can observe that in the case of RNdSK black hole there is a critical value of the Hubble parameter after which the cosmological horizon develops, while in the case of McVittie black hole the apparent horizon is time dependent itself.

\begin{figure}[h!]
\centering
\includegraphics[scale=0.5]{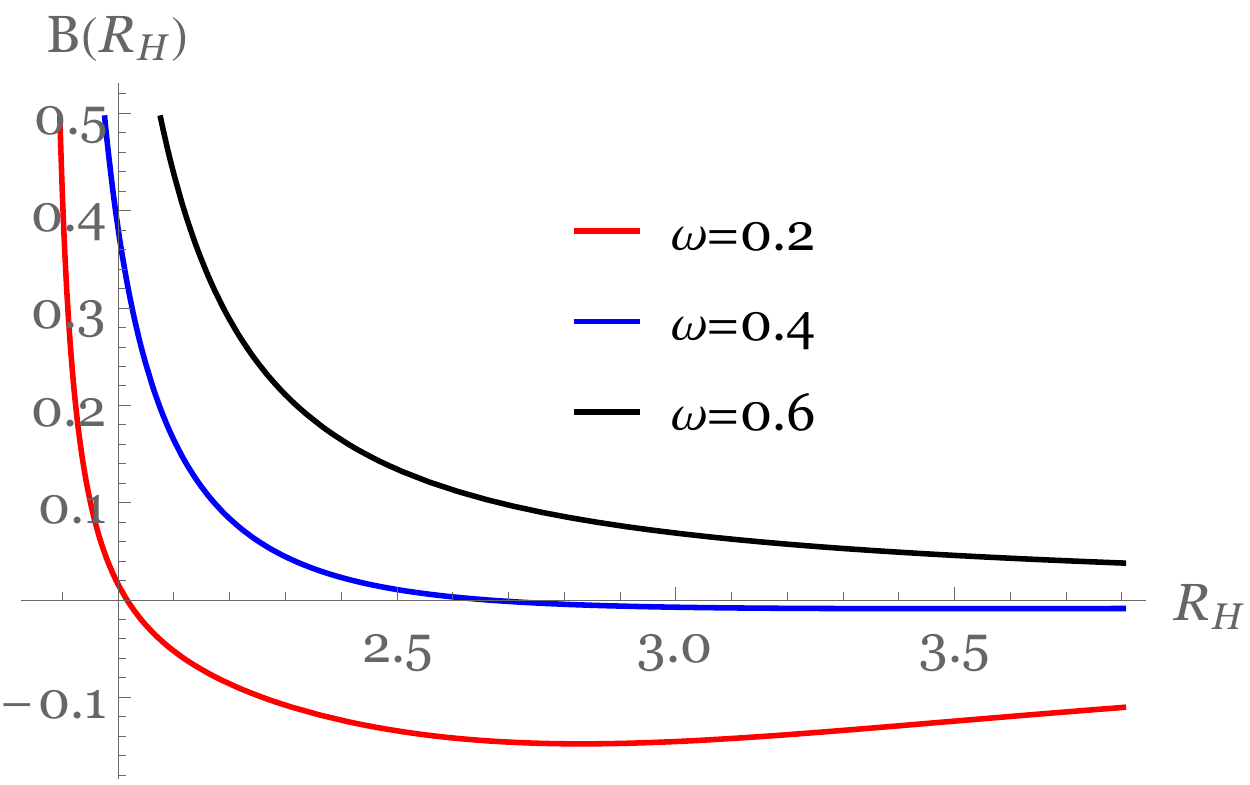}
\caption{Function $B(R_H)$ for $Q=1/2$, $M=1$, $q=1/2$, $\mu=1/2$ for various $\omega$ within the apparent horizon range.}\label{plotB(Rh)versusRhQmiso_matter}
\end{figure}

In  Fig.~\ref{plotB(Rh)versusRhQmiso_matter} we show the superradiance radiation  for various values of the frequency  $\omega$ within the  apparent horizon.  We can see that the value of $B$ is greater than zero always for $\omega=0.6$, while for $\omega=0.4$ we have superradiance only in the range $1.866<R_H<2.661$ of the apparent horizon. As the frequency  $\omega$ is decreasing  the superradiace effect is harder to occur.
\begin{figure}[h!]
\centering
\includegraphics[scale=0.5]{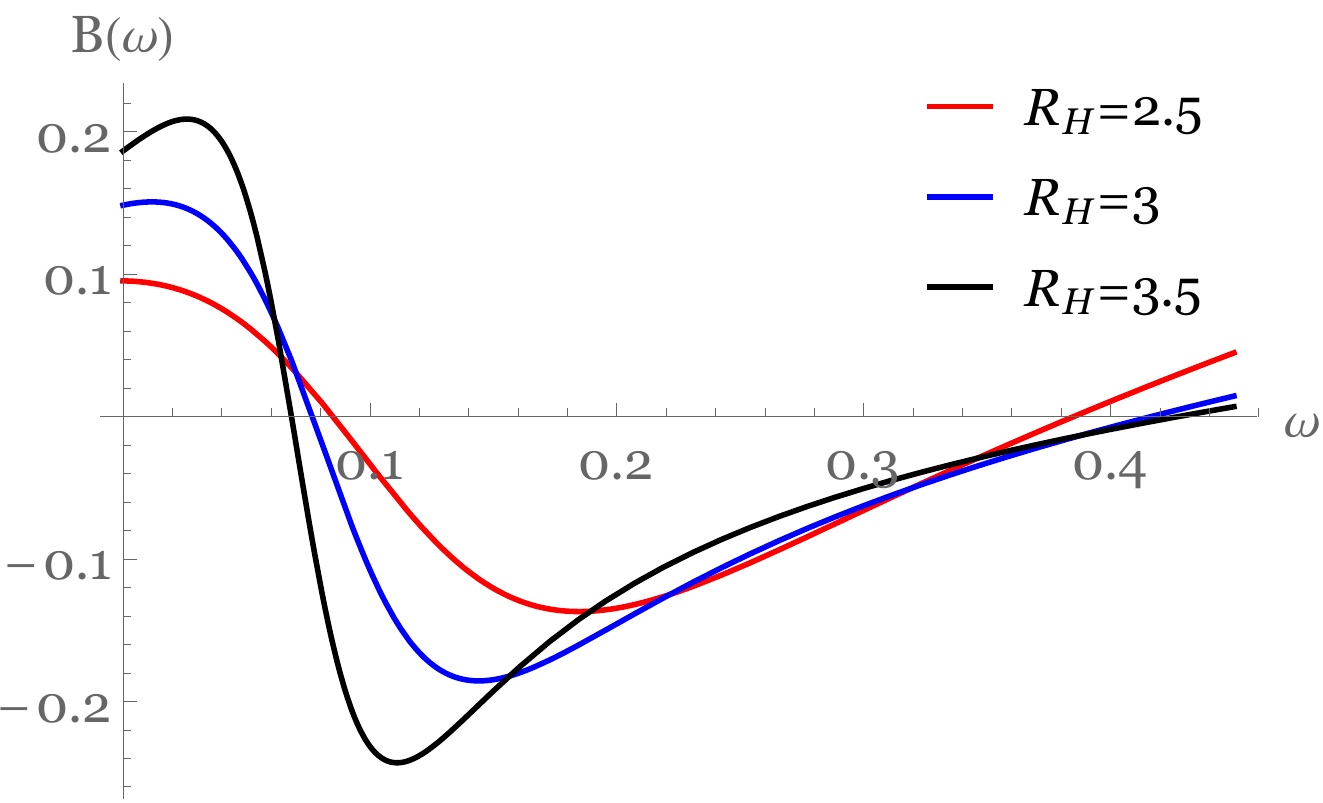}
\caption{Function $B(\omega)$ for $Q=1/2$, $M=1$, $q=1/2$, $\mu=1/2$ for various $R_H$.}\label{plotB(omega)versusomegaQmiso_matter}
\end{figure}

However, as the time is passing and the Universe is expanding the apparent horizon of the McVittie black hole is shrinking and the superradiance radiation can occur more easily for larger range of the frequency $\omega$ as it is shown in  Fig.~\ref{plotB(omega)versusomegaQmiso_matter}.

\begin{figure}[h!]
\centering
\includegraphics[scale=0.5]{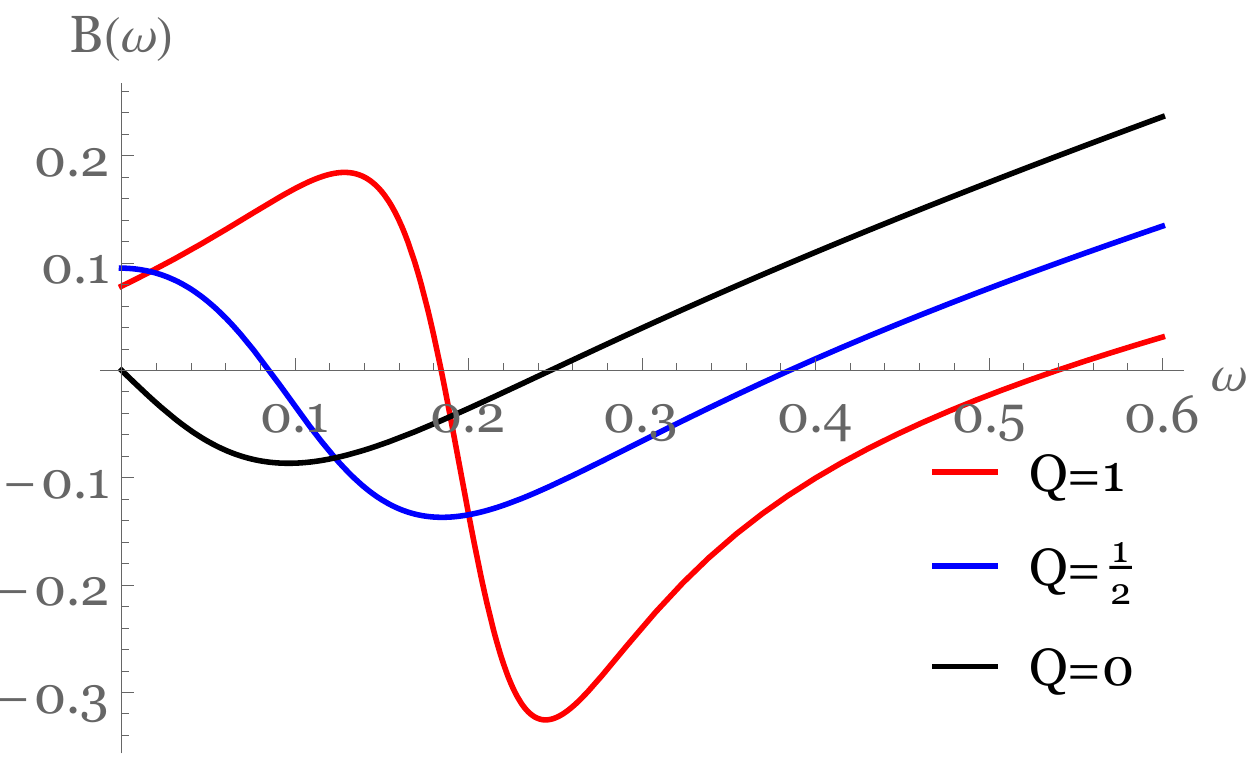}
\caption{Function $B(\omega)$ for $R_H=2.5$, $M=1$, $q=1/2$, $\mu=1/2$ for various $Q$.}\label{plotB(omega)versusomegaQmisoQ1Q0_matter}
\end{figure}

\begin{figure}[h!]
\centering
\includegraphics[scale=0.5]{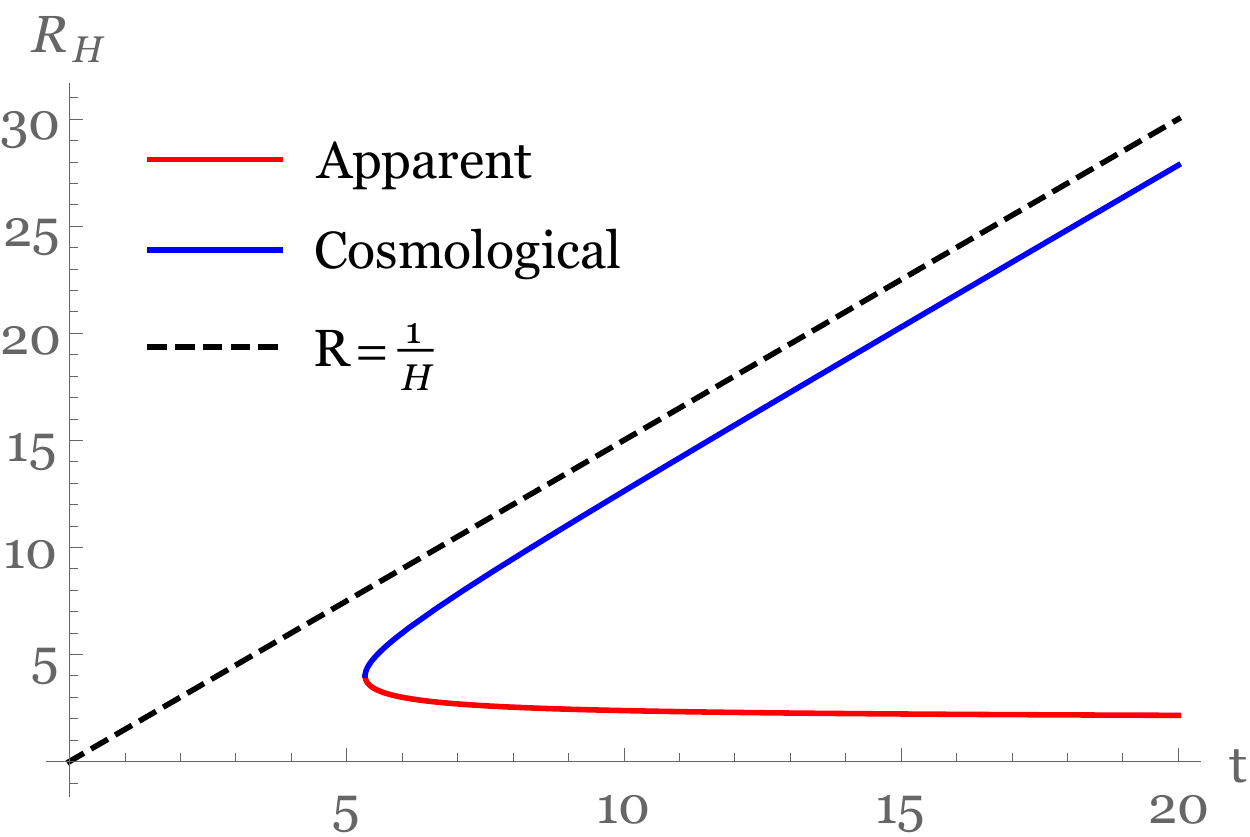}
\caption{Apparent and cosmological horizon for $Q=0$, $M=1$}\label{plotR_HQ0horizons}
\end{figure}

\begin{figure}[h!]
\centering
\includegraphics[scale=0.5]{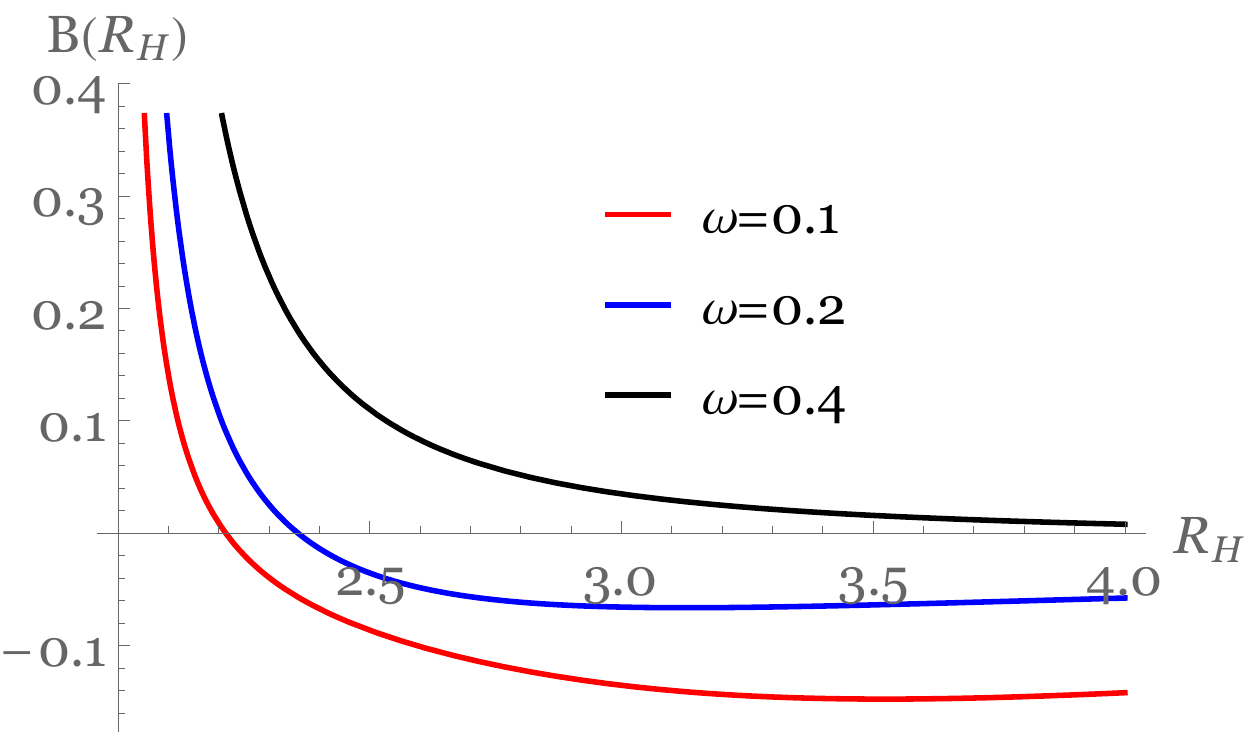}
\caption{Function $B(R_H)$ for $Q=0$, $M=1$, $q=1/2$, $\mu=1/2$ for various $\omega$ within the apparent horizon range.}\label{plotB(Rh)versusRhQ0_matter}
\end{figure}

A very interesting behaviour of the superradiance radiation  is revealed varying the charge $Q$ of the McVittie black hole. In Fig.~\ref{plotB(omega)versusomegaQmisoQ1Q0_matter} we show such a behaviour. We first observe that as the charge  $Q$ increases  the superradiance effect occurs for less values of frequency $\omega$. However, the superradiance radiation occurs for a smaller range of the frequencies $\omega$.
The most interesting effect occurs for $Q=0$. As we can see in Fig.~\ref{plotB(omega)versusomegaQmisoQ1Q0_matter}  we have superradiance radiation even for $Q=0$. In Fig.~\ref{plotR_HQ0horizons} we show the apparent and cosmological horizons of the McVittie black hole without charge. Comparing this figure with  Fig.~\ref{plotR_HQmisohorizons} we observe that the charge does not effect the formation of the horizons.  In Fig.~\ref{plotB(Rh)versusRhQ0_matter} we show the superradiance effect for $Q=0$ for various values of the frequency $\omega$. As can be seen in Fig.~\ref{plotB(Rh)versusRhQ0_matter} we need a large value of the frequency in order to have superradiance radiation. A physical explanation of why we have  superradiance radiation in the neutral McVittie spacetime is that this is happening because the apparent horizon is not static but as the Universe is expanding it is shrinking.
\\

\subsection{Radiation dominated Universe}

We will now study the Klein-Gordon equation \eqref{fullKGR} for a radiation dominated Universe.
 In the radiation dominated Universe from the equation of state $p(t,R)= w \rho(t)$, with $w=1/3$ and from equations (\ref{pieshH}) and (\ref{puknothtaH}) we get
\begin{equation}\label{Hdotradiation}
\dot{H}(t)=-\dfrac{5}{3} H^2(t) \dfrac{\sqrt{Q^2-2 M R+R^2}}{R}~.
\end{equation}
Then equation \eqref{KGLar} admits a solution of separate variables of the form
\begin{equation}\label{solLargeRradiation}
R_{lm}^{rad}(t,R) = R^\alpha T(t)~,
\end{equation}
where $\alpha$ is a complex number. We can also write $R^\alpha = R^{\Re(a)}\; e^{i \Im(\alpha) \ln(R)}$ from which is easy to see that for $\Re(\alpha) < 0 $ the solution goes to zero for $R\to \infty$. Concerning the function of time we find
\begin{equation} \label{eqTradiation}
T(t) \left(\alpha  \dot{H}(t)+\alpha  (\alpha +3) H(t)^2+\mu ^2\right)+(2 \alpha -3) H(t) \dot{T}(t)+\ddot{T}(t) = 0~.
\end{equation}
For a radiation dominated Universe equation \eqref{Hdotradiation} becomes
$
\dot{H}(t) = - \frac{5}{3} H^2(t)$,
with the solution $H(t) = \frac{3}{5 t}$. Substituting this in equation $\eqref{eqTradiation}$ we find
\begin{equation}
T(t)= t^{\frac{1}{5} (-2-3 \alpha)} \mathit{J}_{\frac{2}{5}} (\mu t) C_1 + t^{\frac{1}{5} (-2-3 \alpha)} \mathit{Y}_{\frac{2}{5}} (\mu t) C_2  ~,
\end{equation}
where functions $\mathit{J}_{\frac{2}{5}} (\mu t)$ and $\mathit{Y}_{\frac{2}{5}} (\mu t) $ are the well-known Bessel functions of the first kind $\mathit{J}_n (z)$ and Bessel function of the second kind $\mathit{Y}_n (z)$ respectively.

So the asymptotic solution of the scalar field is
\begin{equation}
\phi_{R \to \infty}^{rad} (t,R,\theta,\phi) = R^\alpha \left(t^{\frac{1}{5} (-2-3 \alpha)} \mathit{J}_{\frac{2}{5}} (\mu t) C_1 + t^{\frac{1}{5} (-2-3 \alpha)} \mathit{Y}_{\frac{2}{5}} (\mu t) C_2\right) Y_{lm}(\theta ,\phi)~.
\end{equation}

At the opposite limit ($R=R_H$) substituting \eqref{apH} and \eqref{Hdotradiation} in the Klein-Gordon equation \eqref{fullKGR}
we find
\begin{align}
&R_{lm}(t,R)\left(\frac{\left((l+1) l+\mu ^2 R^2\right) \left(R (R-2 M)+Q^2\right)}{R^4}+\frac{i q Q \left(3 Q^2-5 M R+R^2 (2+i q Q)\right)}{R^4}\right)\nonumber\\
+ &\frac{\partial R_{lm}(t,R)}{\partial R}\frac{\left(R (R-2 M)+Q^2\right) \left(R (R-8 M)+6 i q Q R^2+7 Q^2\right)}{3 R^5}  \nonumber \\
+& \frac{\partial R_{lm}(t,R)}{\partial t} \frac{2 i q Q R^2+4 Q^2-7 M R+3 R^2}{R^3} \nonumber\\
+& \frac{\partial^2 R_{lm}(t,R)}{\partial R\partial t} \frac{2 \left(R (R-2 M)+Q^2\right)}{R^2} +  \frac{\partial^2 R_{lm}(t,R)}{\partial t^2} = 0~. \label{fullKGRHradiation}
\end{align}
This equation admits a solution of the form
\begin{equation} \label{radialHORradiation}
R_{lm}^{rad}(t,R)=e^{-i \omega t} e^{ \left(A^{rad}+ i B^{rad}\right) R }~,
\end{equation}
where
\begin{align}
A^{rad}=&\dfrac{-3 R_H \left( \left(l(l+1)+\mu^2 R_H^2\right) \left(7 Q^2-8MR_H+R_H^2 \right) + R_H^2 \left(11q^2 Q^2-28 qQR_H \omega+17 R_H^2 \omega^2 \right) \right)}{ \left(\left(7Q^2-8MR_H+R_H^2\right)^2+36 R_H^4 \left(\omega R_H-q Q\right)^2\right)}  ~,\\
B^{rad}=&\dfrac{1}{\left(Q^2-2MR_H+R_H^2\right) \left(\left(7Q^2-8MR_H+R_H^2\right)^2+36 R_H^4\left(\omega R_H-q Q\right)^2\right)}  \nonumber \\
\cdot & \Bigg( -18 R_H^5 \left(\omega R_H-q Q\right)^3  -18 R_H^3 \left(l(l+1)+\mu^2 R_H^2\right)\left(\left(Q^2-2MR_H+R_H^2\right)\left(R_H \omega-qQ\right)\right)\nonumber \\
&+ \Big( -3R_H  \left(7Q^2-8MR_H+R_H^2\right)\left(qQ\left(3Q^2-5MR_H+2 R_H^2\right)-R_H \omega\left(4Q^2-7MR_H+3R_H^2\right)\right) \Bigg) \label{Bradiation}~,
\end{align}
and $\omega$ is the frequency of the mode.

As we discussed in Section \ref{rnostsup}, the solution (\ref{radialHORradiation}) corresponds to the boundary condition (\ref{horsol}) of the static RNdSK black hole and if $B^{rad}>0$ it describes
 an outgoing flux of energy and charge from the charged McVittie black hole.


\begin{figure}[h!]
\centering
\includegraphics[scale=0.5]{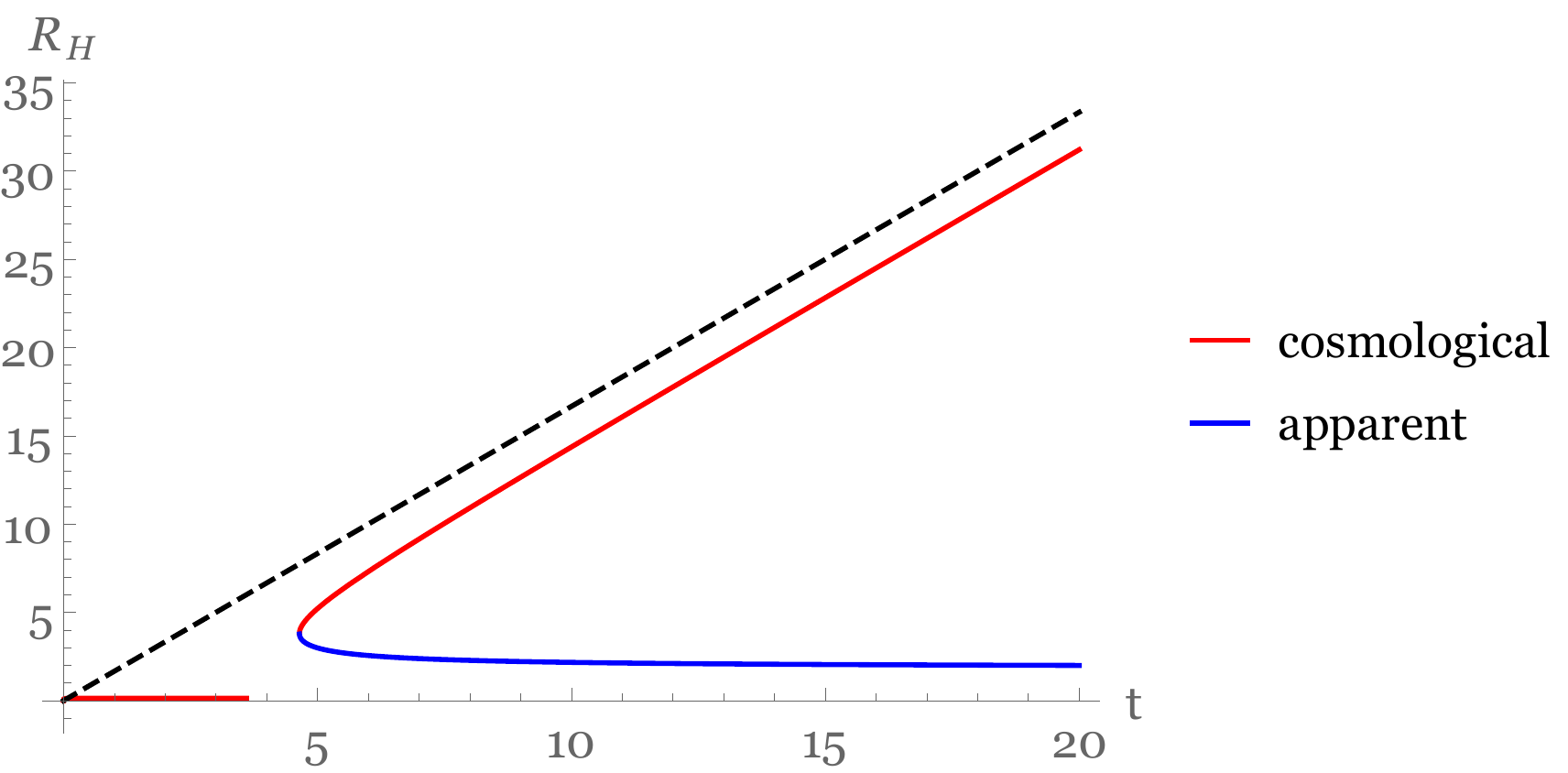}
\caption{Apparent and cosmological horizon for $Q=1/2$, $M=1$}\label{plotR_HQmisohorizonsradiation}
\end{figure}

As in a dust dominated Universe, after a critical time $t^*$ the charged McVittie has an apparent horizon and a cosmological as can be seen in Fig.~\ref{plotR_HQmisohorizonsradiation}. The scattering of a scalar field takes place at the apparent horizon, therefore this critical time can give us an  upper limit to the size of black hole where the superradiance phenomenon occurs. As $t \to \infty$, we see that the apparent horizon shrinks at $M+\sqrt{M^2-Q^2}$ which gives us a lower limit to the $R_H$.

\begin{figure}[h!]
\centering
\includegraphics[scale=0.5]{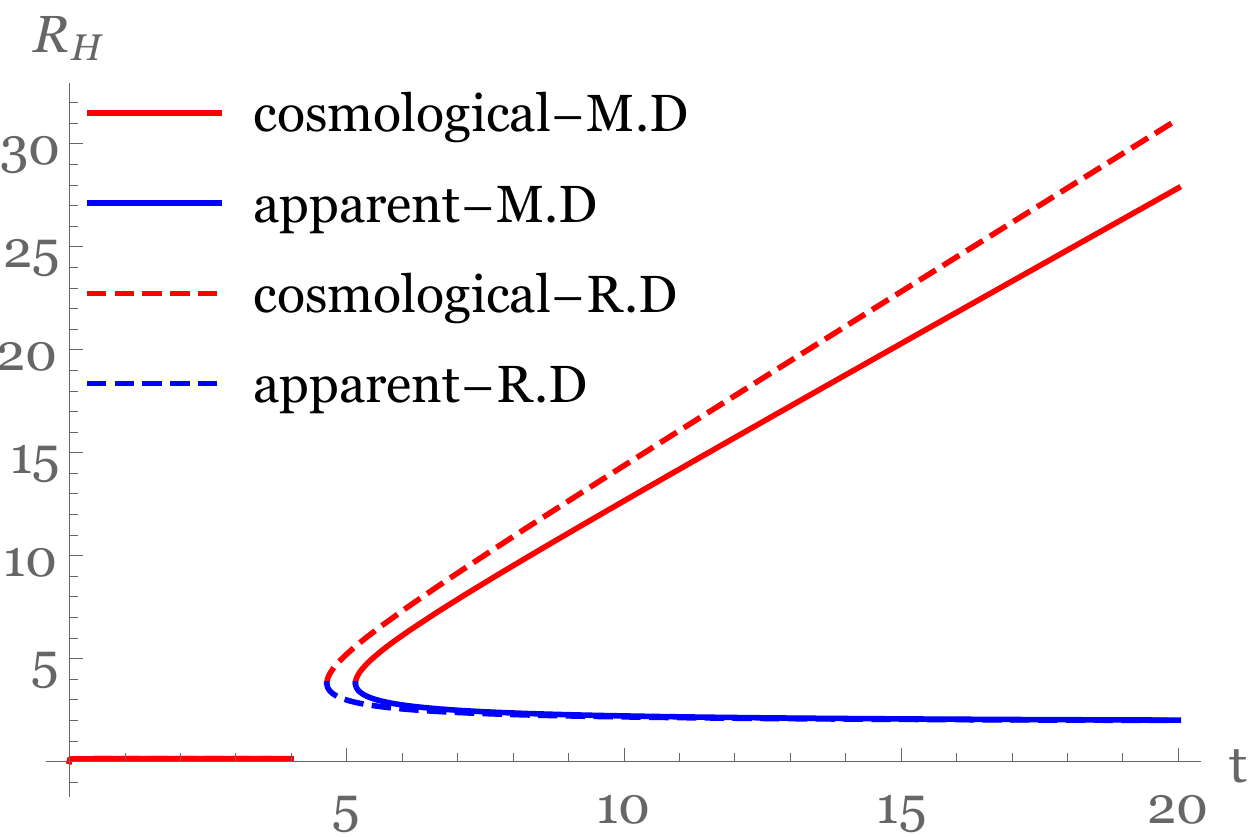}
\caption{Apparent and cosmological horizon for dust and radiation dominated Universe.}\label{plotR_HQmisohorizonsmatter-radiation}
\end{figure}
As we can see in Fig.~\ref{plotR_HQmisohorizonsmatter-radiation}, in a radiation dominated Universe the rate of reduction of the apparent horizon is smaller than in a dust dominated Universe. In the last case the apparent horizon is reduced from an upper value to the lower limit within a shorter range of time. So we expect the parameter $\omega$ to have a similar behaviour as in the previous Section.

\begin{figure}[h!]
\centering
\includegraphics[scale=0.5]{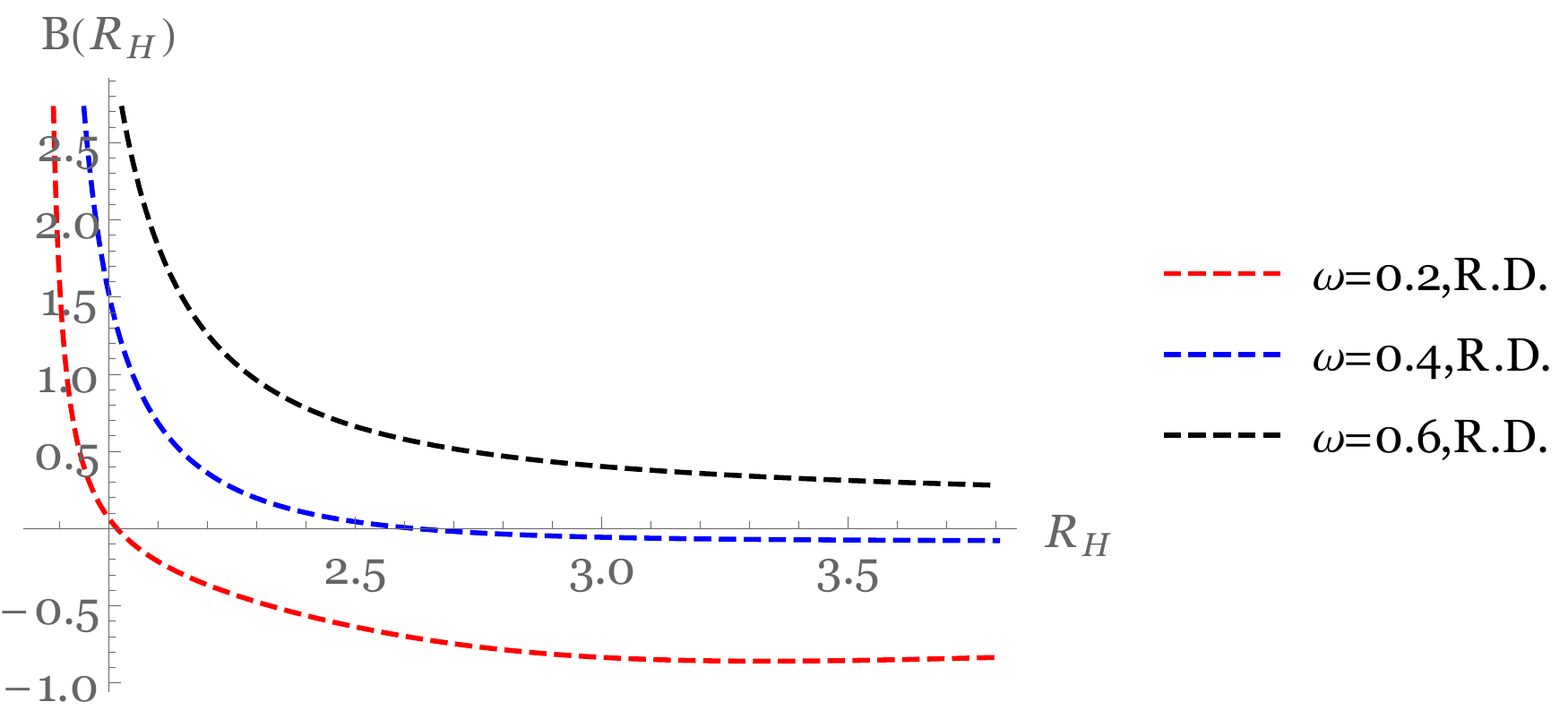}
\caption{Function $B(R_H)$ for $Q=1/2$, $M=1$, $q=1/2$, $\mu=1/2$ for various $\omega$ within the apparent horizon range in a radiation dominated universe.}\label{plotB(Rh)versusRhQmiso_radiation}
\end{figure}

In  Fig.~\ref{plotB(Rh)versusRhQmiso_radiation} we show the superradiance radiation  for various values of the frequency  $\omega$ within the  apparent horizon.  We can see that the value of $B$ is greater than zero always for $\omega=0.6$, while for $\omega=0.4$ we have superradiance only in the range $1.866<R_H<2.621$ (smaller range than in the dust dominated universe) of the apparent horizon. As the frequency  $\omega$ is decreasing  the superradiace effect is harder to occur.
\begin{figure}[h!]
\centering
\includegraphics[scale=0.5]{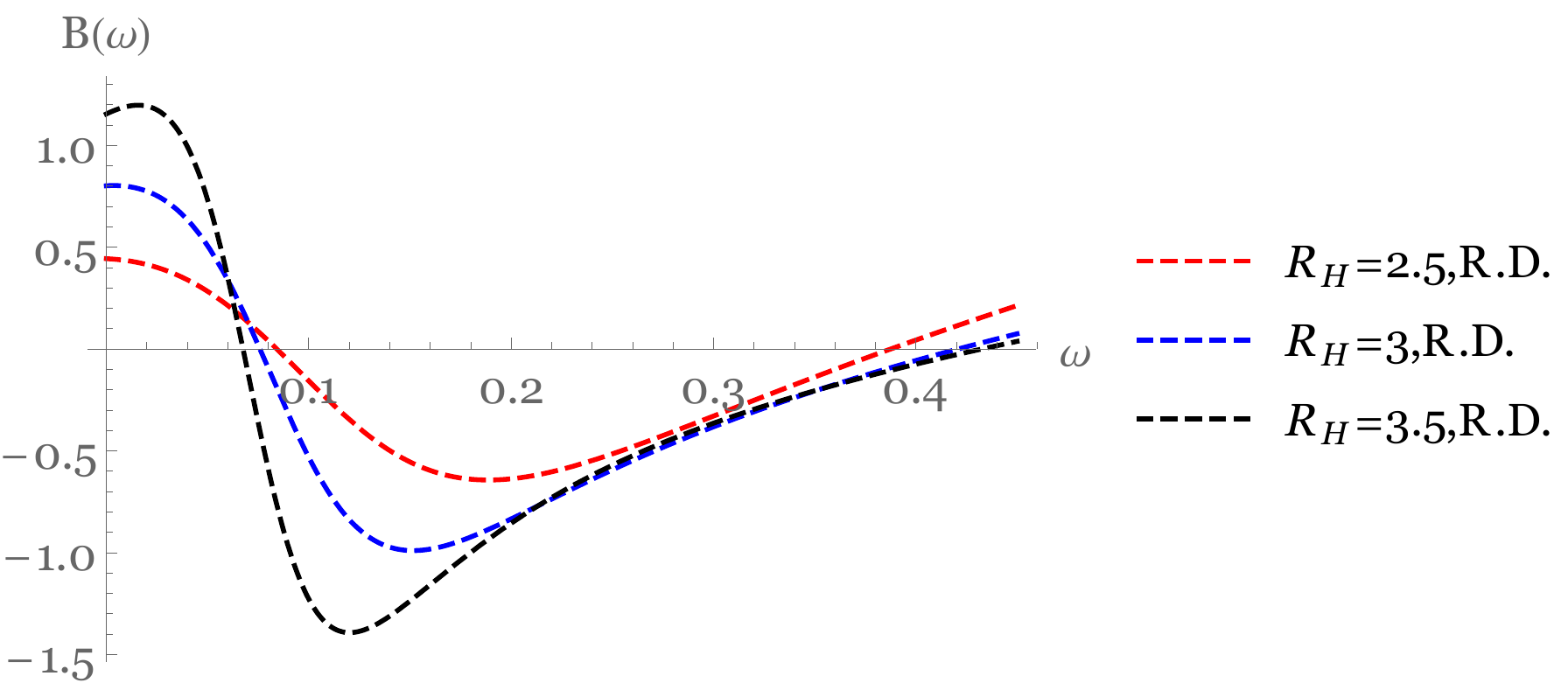}
\caption{Function $B(\omega)$ for $Q=1/2$, $M=1$, $q=1/2$, $\mu=1/2$ for various $R_H$.}\label{plotB(omega)versusomegaQmiso_radiation}
\end{figure}

\begin{figure}[h!]
\centering
\includegraphics[scale=0.5]{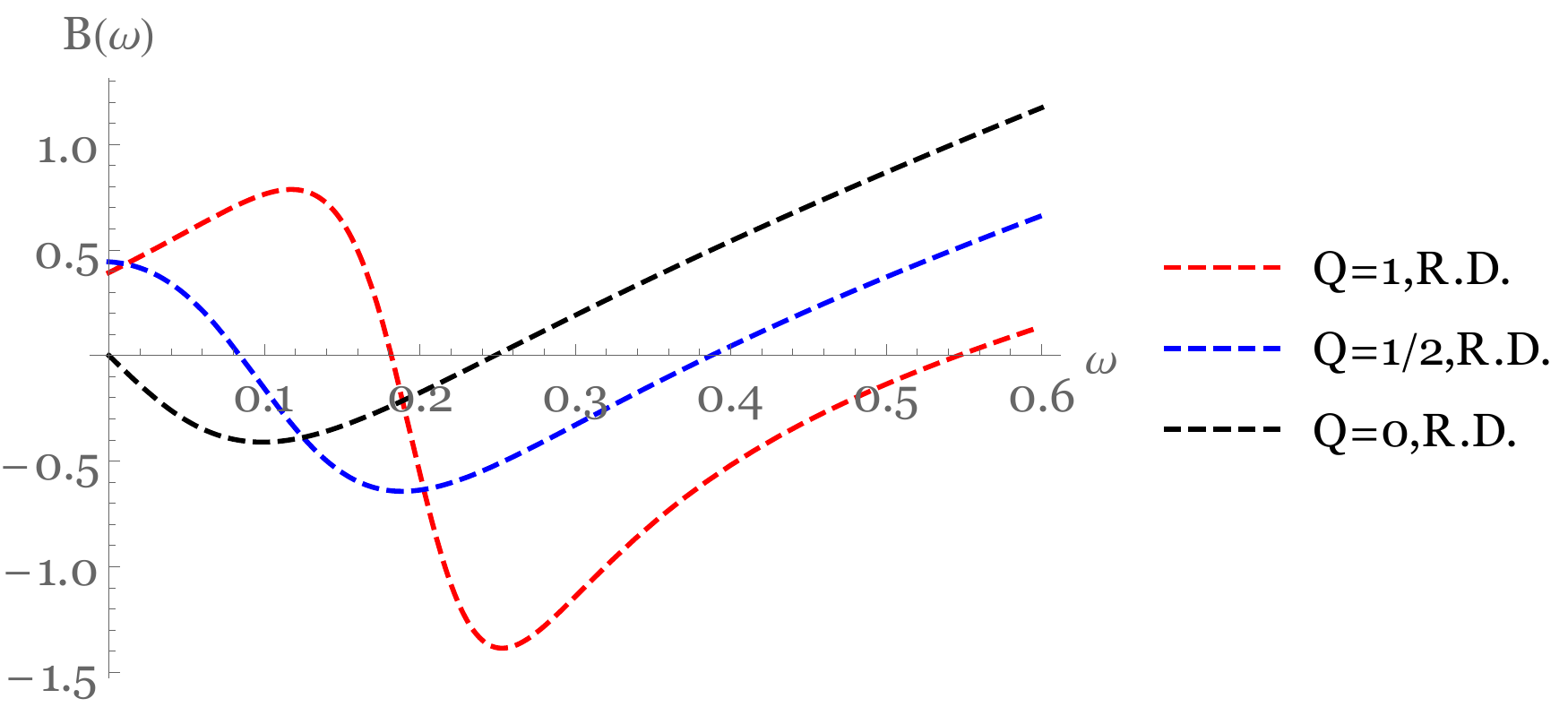}
\caption{Function $B(\omega)$ for $R_H=2.5$, $M=1$, $q=1/2$, $\mu=1/2$ for various $Q$.}\label{plotB(omega)versusomegaQmisoQ1Q0_radiation}
\end{figure}

A similar behaviour to the dust epoch of the superradiance radiation  is revealed varying the charge $Q$ in the radiation dominated epoch.  In Fig.~\ref{plotB(omega)versusomegaQmisoQ1Q0_radiation} we show such a behaviour. We first observe that as the charge  $Q$ increases  the superradiance effect occurs for less values of frequency $\omega$. However, the superradiance radiation occurs for a smaller range of the frequencies $\omega$.
Also in Fig.~{\ref{plotB(Rh)versusRhQ0_radiation} we observe that as in the dust dominated epoch we have superradiance radiation for $Q=0$.

\begin{figure}[h!]
\centering
\includegraphics[scale=0.5]{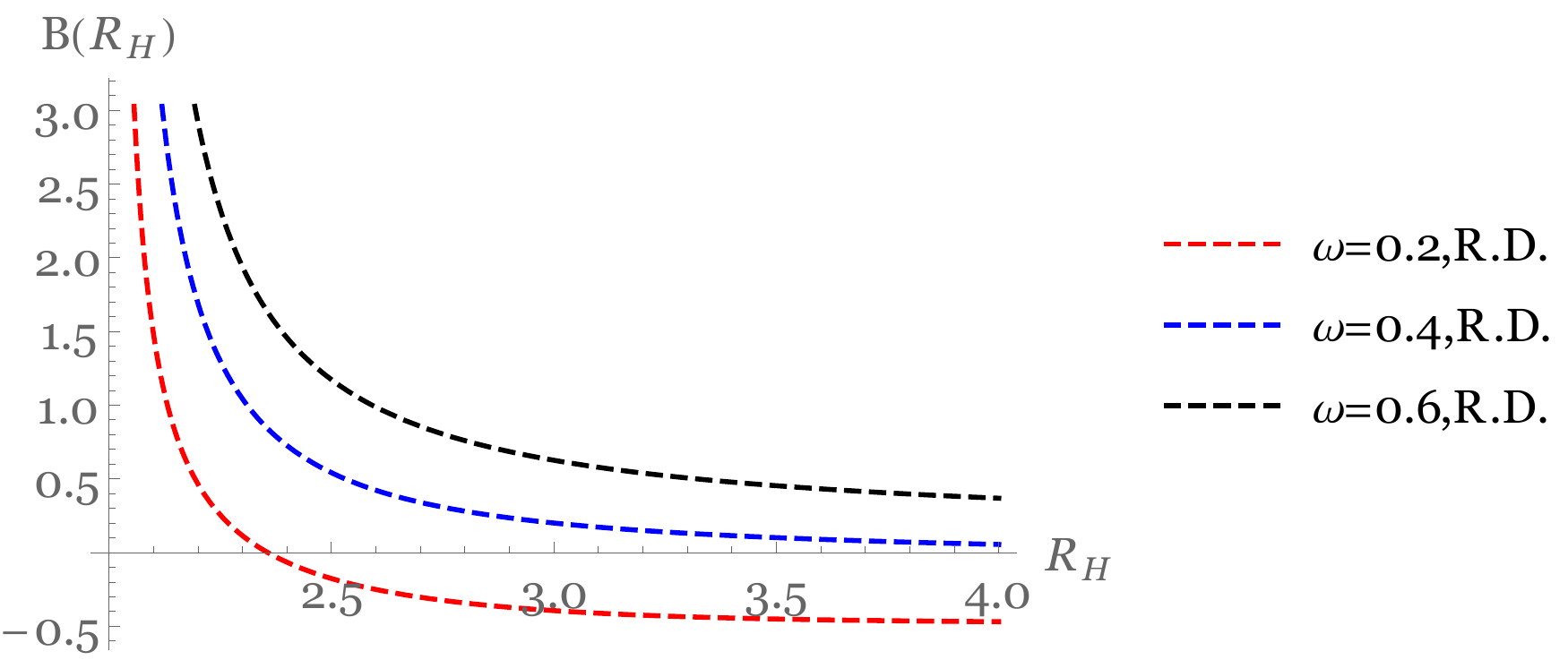}
\caption{Function $B(R_H)$ for $Q=0$, $M=1$, $q=1/2$, $\mu=1/2$ for various $\omega$ within the apparent horizon range.}\label{plotB(Rh)versusRhQ0_radiation}
\end{figure}

\section{Conclusions} \label{conclusion}

We studied  the superradiance effect of a charge black hole  immersed  in an expanding Universe. We considered a test massive charged scalar field scattered off the horizon of the charge McVittie black hole. Because the background is time dependent the radial function of the charged wave should also be time dependent. This leads to a time dependent  Klein-Gordon equation with an explicit dependence on the Hubble parameter $H$ and its derivative $\dot{H}$.
We solved this equation in two different spacial regions, at the asymptotic infinity and at the horizon. We carried out a detailed analysis of the electric energy extracted from the horizon of McVittie black hole in two different epochs of the expansion of the Universe, the dust dominated and radiation dominated epochs.

As the Universe expands  we found that we have extraction of charge and electric energy from the horizon of the charge McVittie black hole. For two epochs of the expansion of Universe, the dust dominated and radiation dominated epochs we studied in details  the superradiance effect for a wide  range of values of the frequency of the scattered wave off the horizon of the charge McVittie black hole. We also provided evidence that
for a neutral McVittie black hole we also have the superradiance effect.  For a range of frequencies of a  scattered wave off the horizon of the neutral  McVittie black hole we  found that there is extraction of energy from its horizon.

 \acknowledgments

 We thank Valerio Faraoni, Pablo Gonz\'{a}lez and Kyriakos Destounis for stimulated discussions.

\begin{appendix}

\section{The Charged McVittie spacetime}

In this Appendix we give a detailed account of the derivation of the charged McVittie black hole. Consider the action
\be
S= \int d^4 x \; \sqrt{-g}\left[\frac{R}{2} + \alpha F^{\mu\nu} F_{\mu\nu} + \mathcal{L}_{\text{fluid}} \right]~.
\ee
Variation with respect to the metric results to the Einstein equations
\be
G_{\mu\nu} = T_{\mu\nu}^{(EM)} + T_{\mu\nu}^{\text{fluid}}~,
\ee
where
\be
 T_{\mu\nu}^{(EM)} = -4 \alpha F^\kappa_\mu F_{\nu\kappa} + \alpha g_{\mu\nu} F_{\kappa\lambda}F^{\kappa\lambda}~,
\ee
and
\be
T_{\mu\nu}^{\text{fluid}} = -\frac{2}{\sqrt{-g}}\frac{\delta \sqrt{-g} \mathcal{L}_\text{fluid}}{\delta g^{\mu\nu}}~.
\ee
A solution to the above equations with the electromagnetic potential as
\begin{equation}
A_\mu=\left\lbrace-\frac{Q}{r a(t) \left(\left(\dfrac{m}{2 r a(t)}+1\right)^2-\dfrac{Q^2}{4 r^2 a(t)^2}\right)},0,0,0\right\rbrace~,
\end{equation}
is the charged McVittie metric
\small
\begin{align}
ds^2=-\frac{\left(1-\dfrac{m^2-Q^2}{4 r^2 a(t)^2}\right)^2}{\left(\left(\dfrac{m}{2 r a(t)}+1\right)^2-\dfrac{Q^2}{4 r^2 a(t)^2}\right)^2} dt^2+a(t)^2 \left(\left(\frac{m}{2 r a(t)}+1\right)^2-\frac{Q^2}{4 r^2 a(t)^2}\right)^2
\left(dr^2 +r^2 d\Omega^2\right)~. \label{mcmetric}
\end{align}
\normalsize
The non-zero components of the Einstein tensor are
\begin{align}
G^t_t &= -\frac{3 \dot{a}^2}{a(t)^2}-\frac{256 Q^2 r^4 a(t)^4}{(2 r a(t)+m-Q)^4 (2 r a(t)+m+Q)^4}~,\\
G^r_r &= -\frac{256 Q^2 r^4 a(t)^4}{(2 r a(t)+m-Q)^4 (2 r a(t)+m+Q)^4} +\frac{\dot{a}(t)^2 (5 (m^2-Q^2) -4 r a(t) (r a(t)-2 m))}{a(t)^2 \left(4 r^2 a(t)^2-m^2+Q^2\right)}~,\\ &\qquad\qquad\qquad \;-\frac{2\ddot{a}(t) (2 r a(t)+m-Q) (2 r a(t)+m+Q)}{a(t) \left(4 r^2 a(t)^2-m^2+Q^2\right)}~,\nonumber\\
G^\theta_\theta = G^\phi_\phi &= \frac{256 Q^2 r^4 a(t)^4}{(2 r a(t)+m-Q)^4 (2 r a(t)+m+Q)^4}+\frac{\dot{a}(t)^2 (5 (m^2-Q^2))-4 r a(t) (r a(t)-2 m))}{a(t)^2 \left(4 r^2 a(t)^2-m^2+Q^2\right)}\\ &\qquad\qquad\qquad \;-\frac{2\ddot{a}(t) (2 r a(t)+m-Q) (2 r a(t)+m+Q)}{a(t) \left(4 r^2 a(t)^2-m^2+Q^2\right)}~.\nonumber
\end{align}
We can rewrite these expressions in a more compact form by using the definition of the  physical areal radius $R$  connected to the comoving radius with the expression \be \label{radius}
R(t,r)= r a(t) \left(\left(\dfrac{m}{2 r a(t)}+1\right)^2-\dfrac{Q^2}{4 r^2 a(t)^2}\right) \quad\Rightarrow\quad 2 a(t) r =  \sqrt{-2 m R+Q^2+R^2}-m+R
\ee
along with the definition of the Hubble parameter
\beno
H(t) = \frac{\dot{a}(t)}{a(t)}, \qquad \ddot{a}(t) = a(t)\left(\dot{H}(t) + H^2(t) \right)~.
\eeno
Then we have
\begin{align}
G^t_t &= -3 H(t)^2-\frac{Q^2}{R^4} \label{e-f1}~,\\
G^r_r &=   -\frac{2 R\dot{H}(t)}{\sqrt{-2 m R+Q^2+R^2}}-3 H(t)^2-\frac{Q^2}{R^4}~, \label{e-f2}\\
G^\theta_\theta = G^\phi_\phi &=   -\frac{2 R\dot{H}(t)}{\sqrt{-2 m R+Q^2+R^2}}-3 H(t)^2+\frac{Q^2}{R^4}~. \label{e-f3}
\end{align}

Assuming the energy momentum tensor of a perfect fluid for ${T^\mu_\nu{}}^\text{fluid}$, that is ${T^\mu_\nu{}}^\text{fluid} =\left( \rho(t)+p(t,r)\right)U^\mu U_\nu + p(t,r)\delta^\mu_\nu$ with $U_\mu = \left(\left(\frac{\left((2 r a(t)+m)^2-Q^2\right)^2}{\left(4 r^2 a(t)^2-m^2+Q^2\right)^2}\right)^{-\frac{1}{2}},0,0,0 \right)$ ,   we find
\begin{align}
{T^t_t{}}^{(EM)} + {T^t_t{}}^{\text{fluid}} &= \frac{2 \alpha  Q^2}{R^4}-\rho (t)~, \\
{T^r_r{}}^{(EM)} + {T^r_r{}}^{\text{fluid}} & = p(t,R)+\frac{2 \alpha  Q^2}{R^4}~,\\
{T^\theta_\theta{}}^{(EM)} + {T^\theta_\theta{}}^{\text{fluid}} = {T^\phi_\phi{}}^{(EM)} + {T^\phi_phi{}}^{\text{fluid}} & = p(t,R)-\frac{2 \alpha  Q^2}{R^4}~.
\end{align}
This means that the total energy momentum tensor of the Einstein equations is indeed consistent with the Einstein tensor components coming from \eqref{mcmetric} if $\alpha = -1/2$ and
\begin{align}
\rho(t) &= 3H^2(t), \label{puknothtaH}\\
p(t,R) &= - \left( \frac{2R \dot{H}(t)}{\sqrt{Q^2-2 mR+R^2}}+3 H(t)^2 \right)~. \label{pieshH}
\end{align}
We can see that the above analysis suggests that we get the neutral McVittie case when $Q=0$ and the FRW background when $Q=m=0$ or when $R\to \infty, (r\to\infty)$. The above expressions can always be written in respect of the comoving radius $r$ as
\begin{align}
\rho (t) &=3\frac{ \dot{a}(t)^2}{a(t)^2}~, \label{densityrho}\\
p(t,r)&= \frac{2 a(t)\ddot{a} (2 r a(t)+m-Q) (2 r a(t)+m+Q)+\dot{a}(t)^2 (4 r a(t) (r a(t)-2 m)-5 (m-Q) (m+Q))}{a(t)^2 \left(4 r^2 a(t)^2-m^2+Q^2~.\right)} \label{momentump}
\end{align}
Observe that the energy density is homogeneous while the  pressure is inhomogeneous. The inhomogeneity of the pressure provides
the necessary  gradient  force to resist  the matter  to fall inside the McVittie black hole horizon in an expanding Universe \cite{Nolan:1999kk,Kaloper:2010ec,Landry:2012nv,Abdalla:2013ara}.

We can finally check that the Bianchi identity
\be
\nabla_\mu G^{\mu\nu} = 0~,
\ee
suggests the energy conservation
\begin{align}
\nabla_\mu T^{\mu\nu} &=
\left\{\frac{((2 r a(t)+m)^2-Q^2) \left(a(t) \rho '(t) ((2 r a(t)+m)^2-Q^2) -3 a'(t) (p(t,r)+\rho (t)) \left(-4 r^2 a(t)^2+m^2-Q^2\right)\right)}{a(t) \left(4 r^2 a(t)^2-m^2+Q^2\right)^2}\right. \nonumber  \\
&\;\; ,\left.\frac{16 r^4 a(t)^2 \left(\frac{4 a(t) (p(t,r)+\rho (t)) \left(4 r a(t) \left(m r a(t)+m^2-Q^2\right)+m (m^2-Q^2)\right)}{4 r^2 a(t)^2-m^2+Q^2}+p^{(0,1)}(t,r) (2 r a(t)+m-Q) (2 r a(t)+m+Q)\right)}{(2 r a(t)+m-Q)^3 (2 r a(t)+m+Q)^3}\right. \nonumber \\
 &\;\ ,0,0\Bigg\rbrace = 0~.
\end{align}
Indeed both components of the above expression are equal to zero by use of equations \eqref{densityrho} and \eqref{momentump} and their derivatives.

The location of the apparent horizons is the solution of
\begin{equation}
\nabla_\mu R \nabla^\mu R = 0~,
\end{equation}
where $R$ is the areal radius. Then we have
\begin{align}
&\left\lbrace\dot{a}^2 r^2\left[\left(1+\dfrac{m}{2ar}\right)^2-\dfrac{Q^2}{4a^2r^2}\right]^4-\left[1-\dfrac{m^2-Q^2}{4a^2r^2}\right]^2\right\rbrace \left[\left(1+\dfrac{m}{2ar}\right)^2-\dfrac{Q^2}{4a^2r^2}\right]^{-2}=0~,
\end{align}
which again using Eq.~\eqref{radius} takes the simpler form
\begin{equation} \label{apHa}
\nabla_\mu R \nabla^\mu R = 0 \quad \Leftrightarrow \quad Q^2-2MR+R^2-R^4H^2(t)=0~.
\end{equation}

We conclude this analysis by writing  the metric (\ref{mcmetric}) using the areal radius $R$
\begin{align} \label{metricRA}
ds^2=-\left(1-\dfrac{2M}{R}+\dfrac{Q^2}{R^2}-R^2H^2\right) dt^2+\dfrac{R^2}{Q^2-2M R+R^2}dR^2
-\dfrac{2RH}{\sqrt{\dfrac{Q^2-2 M R+R^2}{R^2}}} dtdR+R^2d\Omega^2~,
\end{align}
where the electromagnetic potential becomes
\begin{equation}
A_\mu=\left\lbrace-\dfrac{Q}{R},0,0,0\right\rbrace~,
\end{equation}
the tensor of the fluid is ${T^\mu_\nu{}}^\text{fluid} =\left( \rho(t)+p(t,r)\right)U^\mu U_\nu + p(t,r)\delta^\mu_\nu$ with $U_\mu = \left(\dfrac{\sqrt{Q^2-2 M R+R^2}}{R},0,0,0\right)$ and the apparent horizon is the solution of equation \eqref{apHa}.
\end{appendix}

\end{document}